\documentclass[10pt,conference]{IEEEtran}
\usepackage{amsmath,amssymb}
\usepackage{graphicx}
\usepackage{mathrsfs}
\usepackage{tikz}
\usepackage{booktabs}
\usepackage{floatflt}
\usepackage{enumerate}
\usepackage{psfrag}	
\usepackage{array}
\usepackage{multirow,hhline}
\usepackage{exscale}
\usepackage{color}
\usepackage{colortbl}
\usepackage{epsfig}
\newcommand{\mat}[1]{\ensuremath{\mathbf{#1}}}
\renewcommand{\vec}[1]{\ensuremath{\mathbf{#1}}}
\newtheorem{remark}{Remark}
\newtheorem{corollary}{Corollary}
\newtheorem{proposition}{Proposition}

\begin{document}

\IEEEoverridecommandlockouts

% \title{Low $P$ performance of the interference MIMO relay channel}
\title{On the transmit strategy for the interference MIMO relay channel in the low power regime}

\author{
\IEEEauthorblockN{Anas Chaaban and Aydin Sezgin}
\IEEEauthorblockA{Emmy-Noether Research Group on Wireless Networks\\
Institute of Telecommunications and Applied Information Theory\\
Ulm University, 89081, Ulm, Germany\\
Email: {anas.chaaban@uni-ulm.de, aydin.sezgin@uni-ulm.de}}
\thanks{%
This work is supported by the German Research Foundation, Deutsche
Forschungsgemeinschaft (DFG), Germany, under grant SE 1697/3.%
}
}

% \author{\IEEEauthorblockN{Anas Chaaban}%\footnote[]{text}}%\footnotemark}
% % \IEEEautorrefmark{a}
% \IEEEauthorblockA{Institute of Telecommunications\\
% and Applied Information Theory\\
% Ulm University\\
% Ulm, 89081 Germany\\
% email: anas.chaaban@uni-ulm.de}
% \and
% \IEEEauthorblockN{Aydin Sezgin}
% \IEEEauthorblockA{Institute of Telecommunications\\
% and Applied Information Theory\\
% Ulm University\\
% Ulm, 89081 Germany\\
% email: aydin.sezgin@uni-ulm.de}
% \thanks{text}
% % \IEEEcompsocitemizethanks{\IEEEcompsocthanksitem M.
% % Shell is with the Georgia Institute of Technology.
% % \IEEEcompsocthanksitem J. Doe and J. Doe are with An
% % onymous University.}%
% % \thanks{Manuscript received January 20, 2002; revise
% % d January 30, 2002.}
% }

\maketitle

% \footnotetext[1]{a}

\begin{abstract}
This paper studies the interference channel with two transmitters and two receivers in the presence of a MIMO relay in the low transmit power regime. A communication scheme combining block Markov encoding, beamforming, and Willems' backward decoding is used. With this scheme, we get an interference channel with channel gains dependent on the signal power. A power allocation for this scheme is proposed, and the achievable rate region with this power allocation is given. We show that, at low transmit powers, with equal power constraints at the relay and the transmitters, the interference channel with a MIMO relay achieves a sum rate that is linear in the power. This sum rate is determined by the channel setup. We also show that in the presence of abundant power at the relay, the transmit strategy is significantly simplified, and the MAC from the transmitters to the relay forms the bottle neck of the system from the sum rate point of view.
\end{abstract}

\section{Introduction}
The capacity of the interference channel (IC) is a thirty years old problem in network information theory, that is of practical importance as well. When more than one transmitter and receiver want to communicate simultaneously, interference limits their communication. The rate region for the simplest case of two transmitters and two receivers has been thoroughly studied, but the problem remains open for the general case.

Recently, some good achievements have been made in characterizing the degrees of freedom and achievable rate regions of interference networks. It was shown in $\cite{Tse}$, that by using a simple Han-Kobayashi scheme $\cite{HanKob}$, the capacity of a two user interference channel can be achieved to within one bit. For the general case of a $K$-user interference network, it was shown in $\cite{JafCad}$ that the degrees of freedom is given by $K/2$, i.e. the capacity can be well characterized by
\begin{align*}
\frac{K}{2}\log(1+\mathsf{SNR})+o(\mathsf{SNR}),
\end{align*}
where the second term decreases for increasing $\mathsf{SNR}$.
From a practical point of view, it is always interesting to analyze the performance of suboptimal schemes. For instance, in $\cite{ChaSezPau}$, the rate region of a $K$-user interference channel is analyzed for the case in which the interference is treated as noise. The optimality of treating interference as noise for the two-user interference channel has been analyzed in $\cite{VeeAnn,Kha,Kra,BanSezPau}$. Power allocation strategies for the same system have been analyzed in $\cite{Tun08}$. Game-theoretic aspects have been considered in $\cite{JorLar}$. 

Another direction in the study of the IC is the interference relay channel (IRC), where a relay is used to support the communication between transmitters and receivers. This has gained research interest since $\cite{CovGam79}$. Recently, a communication scheme that achieves full degrees of freedom at high $\mathsf{SNR}$ was proposed in $\cite{TanNos08}$ for the interference channel with a MIMO relay (IMRC). In this scheme, the transmitters communicate with the relay in a MAC phase, then the relay broadcasts the received data to the receivers. This is of practical interest, since in practice, the relay does not have knowledge of the transmit signals.

In this paper, we consider the IMRC with the communication scheme proposed in $\cite{TanNos08}$. Namely, this scheme uses superposition block Markov encoding, beamforming, and Willems' backward decoding. In spite of its complexity, this scheme transforms the IMRC to an IC, with channel gains dependent on the signal power, which simplifies the study of the IMRC. In $\cite{TanNos08}$, some power allocation strategies are considered, but these power allocations are not optimal; they are of interest for high transmit power $P$, where they were used to state the degrees of freedom of the system. We extend the study to the low $P$ case, where we study the performance of this scheme, and propose an (approximately) optimal power allocation.

We give the model of the IMRC in section \ref{model}, and describe the communication scheme in section \ref{scheme}. Then we study its performance at low $P$ in section \ref{performance}. A numerical example is included in section \ref{example}. Finally, we conclude with section \ref{conclusion}.

\section{System Model}
\label{model}
\begin{figure}[t]
  \centering{
%      \resizebox{.7\textwidth}{!} {
      \input{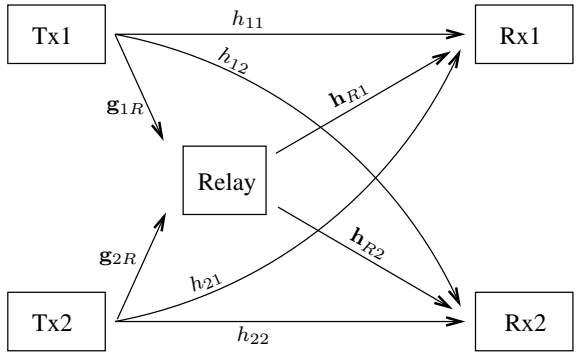}
    }
%    }
%      \includegraphics[width=0.5\textwidth]{Model.pstex}
     \caption{A model for the interference relay channel}
     \label{Model}
\end{figure}

Figure $(\ref{Model})$ shows a model of the IMRC. Each transmitter needs to communicate with its respective receiver, and the relay tries to support this communication. We assume that the transmitters and receivers are equipped with one antenna each, and the relay is equipped with 2 antennas.

We denote by $x_1$, $x_2$, and $\vec{x}_R$ the transmitted signals of transmitter 1, 2 and the relay respectively, and by $y_1$ and $y_2$ the received signals at receivers 1 and 2, respectively. We consider zero mean, unit variance, additive white Gaussian noises at the receivers and the relay denoted as $z_1$, $z_2$, and $\vec{z}_R$. So we can write the input-output relations as:
\begin{eqnarray*}
y_1&=&h_{11}x_1+h_{21}x_2+\vec{h}_{R1}\vec{x}_R+z_1,\\
y_2&=&h_{12}x_1+h_{22}x_2+\vec{h}_{R2}\vec{x}_R+z_2,\\
\vec{y}_R&=&\vec{g}_{1R}x_1+\vec{g}_{2R}x_2+\vec{z}_R,
\end{eqnarray*}
where for $i,j\in\{1,2\}$, $i\neq j$, $h_{ii}$ denotes the direct channel gain from transmitter $i$ to receiver $i$, $h_{ij}$ the cross channel gain from transmitter $i$ to receiver $j$, $\vec{g}_{iR}=[g_{i1}\ g_{i2}]^T$ the channel gain from transmitter $i$ to the relay, and $\vec{h}_{Ri}=[h_{Ri,1}\ h_{Ri,2}]^T$ the channel gain from the relay to receiver $i$. The transmitters have a power constraint $P$, and the relay has a power constraint $P_R$. We assume that the relay operates in full duplex mode, and has global channel knowledge.

\section{Coding Scheme}
\label{scheme}
The coding strategy considered is the one proposed in $\cite{TanNos08}$, and we will briefly explain it in this section. We consider transmission over a period of $B$ blocks, where the sources and the relay send sequences of $B-1$ messages. If a rate pair $(R_1,R_2)$ is achievable in a block, then this scheme achieves a rate pair $(R_1\frac{B-1}{B},R_2\frac{B-1}{B})$, that approaches $(R_1,R_2)$ as ${B\to\infty}$. This coding strategy at the transmitters and the relay is sketched in Table \ref{BMCTable} for the general case, and in the following, we explain it in more details.

\subsection{Encoding at the Sources}
We use super-position block Markov encoding at the sources $\cite{CovGam79}$, i.e.
\begin{eqnarray*}
x_1(b)&=&u_1(b)+u_1'(b),\\
x_2(b)&=&u_2(b)+u_2'(b),\nonumber
\end{eqnarray*}
where for user $i$, $i\in\{1,2\}$, $u_i(b)$ is the codeword of the message of block $b$, with power $p_i$, and $u_i'(b)=\sqrt{\frac{p_i'}{p_i}}u_i(b-1)$, is the codeword of the message of the previous block $b-1$, with power $p_i'$, such that $p_i\in[0,P[$, and $p_i+p_i'=P$. The transmitters use predefined messages $\phi_1$ and $\phi_2$ as the messages of block $0$, i.e. $u_1(0)$ and $u_2(0)$.

\subsection{Decoding and Re-encoding at the Relay}
The relay uses the SDMA scheme described in $\cite[\ Section\ 10.1]{TseVis}$. Assuming that the decoding of messages $u_1(b-1)$ and $u_2(b-1)$ was successful, the relay can subtract them from the received signal, and then decode the messages $u_1(b)$ and $u_2(b)$ using successive interference cancellation, achieving rate constraints given by
\begin{align}
\label{SDMAR1}
R_1&\leq&\log\left(1+\|\vec{g}_{1R}\|^2p_1\right)&=&R_{1}^{MAC},\\
\label{SDMAR2}
R_2&\leq&\log\left(1+\|\vec{g}_{2R}\|^2p_2\right)&=&R_{2}^{MAC},\\
\label{SDMASumRate}
R_1+R_2&\leq&\log\left(\det\left(\mat{I}_2+\mat{G}\mat{K}_p\mat{G}^*\right)\right)&=&R_{sum}^{MAC},
\end{align}
where $\mat{G}=[\vec{g}_{1R}\ \vec{g}_{2R}]$, $\mat{K}_p = diag(p_1,p_2)$, and $\mat{I}_2$ is the $2\times2$ identity matrix.

After decoding, the relay uses multimode beamforming to transmit to the receivers, i.e. the relay constructs the signal
\begin{equation*}
\vec{x}_R(b)=u_{R1}'(b)\vec{t}_1+u_{R2}'(b)\vec{t}_2,
\end{equation*}
where $\vec{t}_1$ and $\vec{t}_2$ are unitary $2\times1$ beamforming vectors. In our approach, $\vec{t}_1$ and $\vec{t}_2$ are chosen such that they reduce interference at the receivers. Let $\rho_1,\rho_2\in[0,1]$ be the power trade-off coefficients at the relay, i.e. the relay splits its power to $\rho_1 P_R$ and $\rho_2 P_R$ for $u_{R1}'(b)$ and $u_{R2}'(b)$ respectively, such that $\rho_1+\rho_2=1$. So
\begin{equation*}
u_{Ri}'(b)=\sqrt{\frac{\rho_i P_R}{p_i'}}u_i'(b), \text{for}\ i\in\{1,2\}.
\end{equation*}

\begin{table*}[ht]
\centering
\begin{tabular}{|c||c|c|c|c|c|c|}
% \hline
% \multicolumn{7}{|c|}{\textbf{Table 1}: Sketch of the superposition block Markov coding scheme.}\\\hline
\hline
block $b$ & 1 & 2 & 3 & \dots & B-1 & B\\\hline
$x_1$&$\left(\phi_1,u_1(1)\right)$&$(u_1(1),u_1(2))$&$(u_1(2),u_1(3))$&\dots&$(u_1(B-2),u_1(B-1))$&$u_1(B-1)$\\\hline
$x_2$&$(\phi_2,u_2(1))$&$(u_2(1),u_2(2))$&$(u_2(2),u_2(3))$&\dots&$(u_2(B-2),u_2(B-1))$&$u_2(B-1)$\\\hline
$\vec{x}_R$&$(\phi_1,\phi_2)$&$(u_1(1),u_2(1))$&$(u_1(2),u_2(2))$&\dots&$(u_1(B-2),u_2(B-2))$&$(u_1(B-1),u_2(B-1))$\\\hline
\end{tabular}
\caption{Sketch of the superposition block Markov coding scheme, here, $\phi_1$ and $\phi_2$ are arbitrary initialization messages known by the transmitters and the relay, and $(x,y)$ means a superposition of $x$ and $y$.}
\label{BMCTable}
\end{table*}

\subsection{Decoding at the destinations}
The received signal at receiver $i$ for block $b$ can be written as
\begin{align*}
y_i(b)=&\ \quad h_{ii}u_i(b)+(h_{ii}+\sqrt{\frac{\rho_i P_R}{p_i'}}\vec{h}_{Ri}^T\vec{t}_i)u_i'(b)\\ &+h_{ji}u_j(b)+(h_{ji}+\sqrt{\frac{\rho_j P_R}{p_j'}}\vec{h}_{Ri}^T\vec{t}_j)u_j'(b)+z_i,\nonumber
\end{align*}
with $i\neq j$, $i,j\in\{1,2\}$. In order to reduce interference, the relay chooses the beamforming vectors $\vec{t}_1$ and $\vec{t}_2$ such that
\begin{eqnarray}
\label{BeamForming}
h_{21}+\sqrt{\frac{\rho_2 P_R}{p_2'}}\vec{h}_{R1}^T\vec{t}_2=0,\\
h_{12}+\sqrt{\frac{\rho_1 P_R}{p_1'}}\vec{h}_{R2}^T\vec{t}_1=0.\nonumber
\end{eqnarray}
Let us denote by $\vec{t}_{10}$ and $\vec{t}_{20}$ the vectors $\sqrt{\frac{\rho_1 P_R}{p_1'}}\vec{t}_1$ and $\sqrt{\frac{\rho_2 P_R}{p_2'}}\vec{t}_2$ respectively. Since $\vec{t}_1$ and $\vec{t}_2$ are unitary, it follows
\begin{eqnarray}
\label{Unitary}
\|\vec{t}_{10}\|^2&=&\frac{\rho_1 P_R}{p_1'},\\
\|\vec{t}_{20}\|^2&=&\frac{\rho_2 P_R}{p_2'}.\nonumber
\end{eqnarray}
With $(\ref{BeamForming})$ and $(\ref{Unitary})$, we get a system of two equations with two unknowns for each of the beamforming vectors. Notice that these equations do not have a unique solution. Equation $(\ref{BeamForming})$ tells us that the components of $\vec{t}_{i0}$ are linear with respect to each other, while $(\ref{Unitary})$ tells us that the beamforming vector lies on a circle, leading to two solutions. Solving for $\vec{t}_{10}$ and $\vec{t}_{20}$, we get for $i\neq j$, $i,j\in\{1,2\}$,
\begin{equation}
\label{BeamFormingVector}
\vec{t}_{i0}= \left[ \begin{array}{c}
\frac{1}{\|\vec{h}_{Rj}\|^2}T_{i0} \\
-\frac{h_{ij}}{h_{Rj,2}}-\frac{1}{\|\vec{h}_{Rj}\|^2}\frac{h_{Rj,1}}{h_{Rj,2}}T_{i0} \end{array} \right],
\end{equation}
where
\begin{equation}
\label{sqT}
T_{i0}=n_i h_{Rj,2}\sqrt{-h_{ij}^2+\|\vec{h}_{Rj}\|^2\frac{\rho_i P_R}{P-p_i}}-h_{ij}h_{Rj,1},
\end{equation}
with $n_i\in\{-1,1\}$. This gives unitary $\vec{t}_1$ and $\vec{t}_2$, and satisfies $(\ref{BeamForming})$. This choice of $\vec{t}_{i0}$ reduces interference seen by the receivers, so then we can express $y_i(b)$ as
\begin{equation*}
y_i(b)=h_{ii}u_i(b)+(h_{ii}+\vec{h}_{Ri}^T\vec{t}_{i0})u_i'(b)+h_{ji}u_j(b)+z_i.
\end{equation*}
Now, the receivers can use Willems' backward decoding $\cite{Wil82}$ to decode their signals. Starting from block $B$, receivers 1 and 2 have interference free signals and can decode $u_1(B-1)$ and $u_2(B-1)$ respectively. Then, in each block $b$, the receivers subtract the already known signals $u_1(b)$ and $u_2(b)$ from their received signals before attempting to decode $u_1(b-1)$ and $u_2(b-1)$. Now we can express $y_i(b)$ as
\begin{equation}
\label{InOut}
y_i(b)=(h_{ii}+\vec{h}_{Ri}^T\vec{t}_{i0})u_i'(b)+h_{ji}u_j(b)+z_i.
\end{equation}
As a result, the interference relay channel transforms into an IC. To simplify the notation, we will use $f_{11}, f_{12}, f_{21},$ and $f_{22}$ to denote the new channel coefficients:
\begin{eqnarray}
\label{CandEofp1p2}
f_{ii}=h_{ii}+\vec{h}_{Ri}^T\vec{t}_{i0}, &\ f_{ij}=h_{ij}.
\end{eqnarray}
Now we can write the obtained IC input-output equations $(\ref{InOut})$ as
\begin{eqnarray*}
y_i(b)=f_{ii}u_i'(b)+f_{ji}u_j(b)+z_i,
\end{eqnarray*}
where $f_{ii}$ and $f_{ji}$ depend on the channel coefficients, $p_i$, $P$, $P_R$ and $\rho_i$.
\section{Performance at low transmit power $P$}
\label{performance}
We aim in this section to analyze the performance of the given scheme at low transmit power $P$. Denote the optimal power allocation at the transmitters for a fixed power allocation $\rho_i$ as $\tilde{p}_1$ and $\tilde{p}_2$, and denote the rate region achieved by this power allocation as $\mathcal{R}_\rho$. Then we have the following proposition.
\begin{proposition}
The rate region $\mathcal{R}$ of the IMRC with the considered scheme, at low $P$ is given by
\begin{equation*}
\mathcal{R}=ch\left(\bigcup_{\rho\in[0,1]}\mathcal{R_\rho}\right),
\end{equation*}
where $ch(\mathcal{S})$ denotes the convex hull of $\mathcal{S}$.
\end{proposition}
\subsection{Treating interference as noise}
 Let us assume for the moment being, that we fix a choice of $\vec{t}_{10}$ and $\vec{t}_{20}$, and we consider a fixed power allocation at the relay, i.e. fixed $n_i$ and $\rho_i$. Since $p_i<P$, we can approximate $f_{11}$ and $f_{22}$ as linear functions of $p_1$ and $p_2$ respectively as follows (see details in appendix $\ref{FirstApprox}$)
\begin{eqnarray}
\label{CApprox}
f_{ii}\approx \mu_{ii}+\nu_{ii}\frac{p_i}{P},
\end{eqnarray}
where we drop the arguments of $f_{ii}^{(0)}$, and $f_{ii}^{(1)}$ for readability. This approximation is  needed for solving our optimization problem, due to the fact that the argument of the square root in $(\ref{sqT})$ is not concave in $p_i$, and hence can not be optimized using standard convex optimization tools (e.g. $\cite{Boyd}$).

The receivers in the obtained IC treat interference as noise, resulting in rates bounded by
\begin{eqnarray}
\label{TreatingInterferenceAsNoiseRates1}
R_1\leq\log\left(1+\frac{\|f_{11}\|^2(P-p_1)}{1+\|f_{21}\|^2p_2}\right)=R_{1}^{IC},\\
\label{TreatingInterferenceAsNoiseRates2}
R_2\leq\log\left(1+\frac{\|f_{22}\|^2(P-p_2)}{1+\|f_{12}\|^2p_1}\right)=R_{2}^{IC}.
\end{eqnarray}
\subsection{Power allocation at low $P$ for sum rate maximization}
Up to this point, the expressions are not low-$P$-specific. From this point on, we restrict ourself to low $P$. We still consider fixed $n_i$ and $\rho_i$.
Let us write the rate region for this scenario as
\begin{eqnarray*}
R_1&\leq&\min(R_{1}^{MAC},R_{1}^{IC}),\nonumber\\
R_2&\leq&\min(R_{2}^{MAC},R_{2}^{IC}),\nonumber\\
R_1+R_2&\leq&R_{sum}^{MAC}.\nonumber
\end{eqnarray*}
It is required to find powers $p_i$ that maximize this region. In the following proposition, we will specify this rate region at low $P$ for fixed arbitrary $\rho_i$ and $n_i$, the proof is shown in Appendix $\ref{SecondApprox}$.
\begin{proposition}
\label{RateRegion}
The rate region of the IMRC, with the coding scheme described in section \ref{scheme}, with fixed $n_i$ and $\rho_i$ can be approximated at low $P$ as
\begin{eqnarray}
\label{RateRegionBounds}
R_1&\leq&\frac{\|\vec{g}_{1R}\|^2\hat{p}_1}{\ln{2}},\\
R_2&\leq&\frac{\|\vec{g}_{2R}\|^2\hat{p}_2}{\ln{2}},\nonumber
% R_1+R_2&\leq&\frac{\|\vec{g}_{1R}\|^2 \hat{p}_1+\|\vec{g}_{2R}\|^2 \hat{p}_2}{\ln{2}},\nonumber
\end{eqnarray}
where 
$$\hat{p}_1=\frac{\lambda_1+\sqrt{\lambda_1^2+8\|\mu_{11}\|^2\Re(\mu_{11}\nu_{11}^*)}}{4\Re(\mu_{11}\nu_{11}^*)}P,$$
$$\hat{p}_2=\frac{\lambda_2+\sqrt{\lambda_2^2+8\|\mu_{22}\|^2\Re(\mu_{22}\nu_{22}^*)}}{4\Re(\mu_{22}\nu_{22}^*)}P,$$ $\lambda_1=2\Re(\mu_{11}\nu_{11}^*)-\|\mu_{11}\|^2-\|\vec{g}_{1R}\|^2$, and $\lambda_2=2\Re(\mu_{22}\nu_{22}^*)-\|\mu_{22}\|^2-\|\vec{g}_{2R}\|^2$.
\end{proposition}
Notice that the rate bounds in proposition $\ref{RateRegion}$ are linear in $\hat{p}_1$ and $\hat{p}_2$, which are functions of $n_1$ and $n_2$, so we have the following corollary.
\begin{corollary}
\label{FixedRhoPower}
The rate region in proposition $\ref{RateRegion}$ is maximized for a fixed arbitrary $\rho_i$ by choosing powers
\begin{eqnarray*}
\tilde{p}_1&=&\max_{n_1\in\{-1,1\}} \hat{p}_1,\nonumber\\
\tilde{p}_2&=&\max_{n_2\in\{-1,1\}} \hat{p}_2.\nonumber
\end{eqnarray*}
Plugging these powers in $(\ref{RateRegionBounds})$, we get the region $\mathcal{R}_\rho$.
\end{corollary}
\subsection{Special Case: $P_R\gg P$}
In this subsection, we introduce a special case, which has the advantage of significantly simplifying the transmit strategy. Namely, we consider the case of abundant power at the relay, i.e. $P_R\gg P$. In this case, we can approximate $\vec{t}_{i0}$ as
\begin{align*}
\vec{t}_{i0}\approx \left[ \begin{array}{c}
\frac{n_ih_{Rj,2}}{\|\vec{h}_{Rj}\|}\sqrt{\frac{\rho_1P_R}{P-p_i}} \\
-\frac{n_ih_{Rj,2}}{\|\vec{h}_{Rj}\|}\sqrt{\frac{\rho_1P_R}{P-p_i}} \end{array} \right],
\end{align*}
where $n_i\in\{-1,1\}$. It follows that the coefficients of the IC become
\begin{align*}
f_{11}\approx n_1\frac{\det(\mat{H})}{\|\vec{h}_{R2}\|}\sqrt{\frac{\rho_1P_R}{P-p_1}},\\
f_{22}\approx n_2\frac{\det(\mat{H})}{\|\vec{h}_{R1}\|}\sqrt{\frac{\rho_2P_R}{P-p_2}}.
\end{align*}
Substituting in $(\ref{TreatingInterferenceAsNoiseRates1})$ and $(\ref{TreatingInterferenceAsNoiseRates2})$, we get the following for $R_1^{IC}$ and $R_2^{IC}$:
\begin{align*}
R_1^{IC}\approx\log\left(1+\frac{\det^2(\mat{H})\rho_1P_R}{\|\vec{h}_{R2}\|^2(1+\|f_{21}\|^2p_2)}\right)=R_{1}^{AP},\\
R_2^{IC}\approx\log\left(1+\frac{\det^2(\mat{H})\rho_2P_R}{\|\vec{h}_{R1}\|^2(1+\|f_{12}\|^2p_1)}\right)=R_{2}^{AP}.
\end{align*}
If $P_R$ is high enough, then the rates with abundant relay power $R_{1}^{AP}$ and $R_{2}^{AP}$ are greater than the rates at the MAC side of the IMRC $R_{1}^{MAC}$ and $R_{2}^{MAC}$ respectively for all $p_1$ and $p_2$. Consequently, the sum rate is determined by the MAC side of the IMRC, i.e. by $R_1^{MAC}$ and $R_{2}^{MAC}$, and the optimal power allocation for maximizing the sum rate in this case is $p_1=p_2=P$.
\begin{remark}
The expressions in section $\ref{scheme}$ are defined for $p_i\in[0,P[$, however, they can be easily modified to include $p_i=P$.
\end{remark}
As a result, at high $P_R$, the transmitters do not need to use super-position block Markov encoding. Each transmitter sends $u_i(b)$ in block $b$, the relay decodes $u_i(b)$, and then sends them delayed at the next block $b+1$ while still using multimodal beamforming. In this case, we achieve $R_i^{MAC}=\log(1+\|\vec{g}_{iR}\|^2P)$.
\section{Numerical Example}
\label{example}
Consider the channel with parameters
\begin{eqnarray*}
h_{11}=h_{22}=1.2&,&\ h_{12}=h_{21}=0.5,\nonumber\\
\vec{g}_{1R}=[0.6\ 1.2]^T&,&\ \vec{g}_{2R}=[1\ 0.5]^T,\nonumber\\
\vec{h}_{R1}=[0.5\ 1]^T&,&\ \vec{h}_{R2}=[1\ 2]^T,\nonumber\\
\end{eqnarray*}
and assume $P_R=P=0.1$.
\begin{figure}[]
     \centering
{
\psfragscanon
\psfrag{xlabel}[][]{\footnotesize $p_1/P$}
\psfrag{ylabel}[][]{\footnotesize Components of $t_{10}$}
\tiny
\psfrag{leg1}[l][]{$\vec{t}_{10}$, $1^{st}$ component, Exact}
\psfrag{leg2}[l][]{$\vec{t}_{10}$, $2^{nd}$ component, Exact}
\psfrag{leg3}[l][]{$\vec{t}_{10}$, $1^{st}$ component, Approximate}
\psfrag{leg4}[l][]{$\vec{t}_{10}$, $2^{nd}$ component, Approximate}
          \includegraphics[width=.4\textwidth]{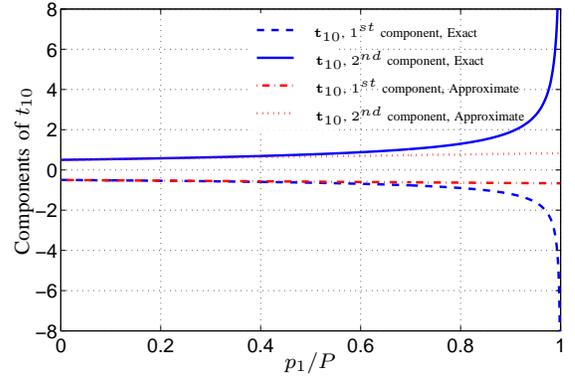}}
     \caption{Components of the beamforming vector $\vec{t}_{10}$ plotted as a function of $p_1$ for $P=0.1$. The plot also shows our approximation for the components of $\vec{t}_{10}$.}
\label{t1}
\end{figure}
\begin{figure}[]
\centering
{
\psfragscanon
\psfrag{xlabel}[][]{\footnotesize $p_1/P$}
\tiny
\psfrag{tag1}[l][]{$r_{1}^{MAC}$}
\psfrag{tag2}[l][]{$r_{1}^{IC}$}
\psfrag{tag3}[l][]{$R_{1}^{MAC}$}
\psfrag{tag4}[l][]{$R_{1}^{IC}$}
\psfrag{p1hat}[c][]{$\hat{p}_1$ delivering max. $R_1$}
          \includegraphics[width=.4\textwidth]{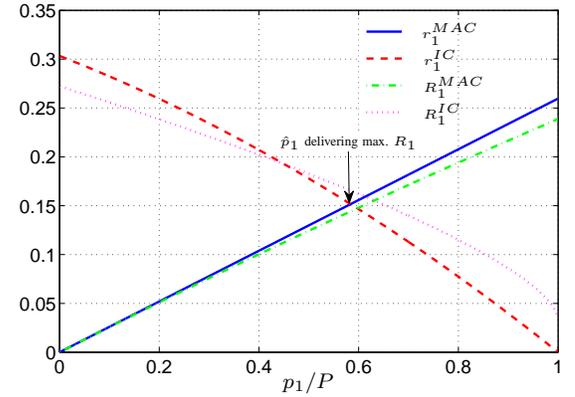}}\\
     \vspace{0cm}
%     \hspace{.1in}
      \caption{Plots of the functions $r_{1}^{MAC}$ and $r_{1}^{IC}$, showing their intersection at the power that delivers optimal $R_1$. It also shows plots of the exact expressions $R_{1}^{MAC}$ and $R_{1}^{IC}$ (for arbitrary small $p_2<P$).}
     \label{Rofp1}
\end{figure}
For equal power split at the relay, i.e. $\rho=0.5$, the components of beamforming vector $\vec{t}_1$, and its approximation are shown in figure $(\ref{t1})$.
Figure $(\ref{Rofp1})$ shows a plot of $r_{1}^{MAC}$ and $r_{1}^{IC}$. Notice that in this example, if we choose $\hat{p}_1=0.0583$, then we maximize $\min(r_{1}^{MAC},r_{1}^{IC})$, allowing us to achieve maximum $R_1$.

\begin{figure}[]
     \centering
\psfragscanon
\psfrag{xlabel}[][]{\footnotesize $P(dB)$}
\psfrag{ylabel}[][]{\footnotesize $\hat{p}_1/P$}
\tiny
\psfrag{num}[l][]{Exhaustive search solution}
\psfrag{app}[l][]{Low $P$ approximation}
     \includegraphics[width=0.4\textwidth]{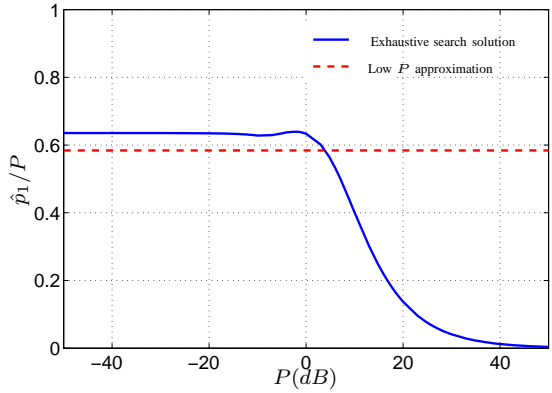}
	\caption{Exhaustive search and approximate solution of the power allocation problem for $p_1$ (normalized to $P$).}
     \label{ExactvsApprox}
\end{figure}

\begin{figure}[]
     \centering
\psfragscanon
\psfrag{x}[][]{\footnotesize $P(dB)$}
\psfrag{y}[][]{\footnotesize Normalized Sum Rate}
\tiny
\psfrag{num}[l][]{Exhaustive search solution}
\psfrag{app}[l][]{Low $P$ approximate solution}
\psfrag{equ}[l][]{$p_1=p_2=P/2$}
\psfrag{squ}[l][]{$p_1=p_2=\sqrt{P}$}
     \includegraphics[width=0.4\textwidth]{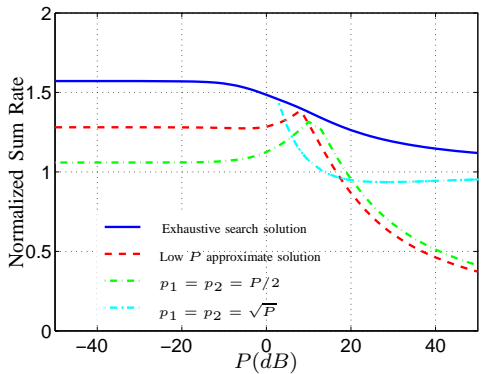}
	\caption{Normalized sum rates for different power allocation strategies.}
     \label{SumRateAll}
\end{figure}

Figure $(\ref{ExactvsApprox})$ shows the exhaustive search (numerical) solution of the power allocation problem ($\hat{p}_1$) and the approximate solution, both normalized to $P$.
Finally, in figure $(\ref{SumRateAll})$, we show the sum rates (normalized to $\log(1+\|h_{ii}\|^2P)$) for four different power allocations: 
\begin{itemize}
\item Optimal power allocation for maximum sum rate (exhaustive search),
\item Approximate power allocation as in corollary $\ref{FixedRhoPower}$,
\item Equal power allocation with $p_1=p_2=P/2$, and
\item Equal power allocation with $p_1=p_2=\sqrt{P}$ for $P\geq1$.
\end{itemize}
Notice that at low $P$, our approximation (dashed line) is close to the maximal sum rate, and that it is constant in that region. Notice also that the maximum sum rate approaches one for large $P$. The power allocation $p_1=p_2=\sqrt{P}$ and $p_1=p_2=P/2$ give a normalized sum rate approaching zero and one respectively at high $P$ which confirms results in $\cite{TanNos08}$.
\section{Conclusion}
\label{conclusion}
As a result of this work, we have obtained an approximation for the optimal power allocation, that maximizes the sum rate for the given scheme. If we consider the special case of $P_R=P$, then we obtain a sum rate that is linear in $P$ at low $P$. It follows that the normalized sum rate is a constant at low $P$, given by the channel parameters and the power split at the relay. 

Using super-position block Markov encoding at the sources, beamforming at the relay, and Willems' backward decoding at the receivers, the IMRC transforms into an IC. We have given the channel gains of this IC, as functions of the parameters of the system, including the powers.

Of practical interest is the case where the relay power is much greater than the transmit power. In this case, we have shown that the encoding at the transmitters becomes simpler, since there is no need to perform super-position block Markov encoding. Furthermore, the MAC from the transmitters to the relay forms the bottle neck for the system from the sum rate point of view in this case.

Given the obtained IC, the question of the optimality of treating interference as noise at the receivers arises. It would be interesting to find conditions on this channel that allow us to optimally treat interference as noise. This work can also be extended to the high power regime, where an optimal power allocation that maximizes the sum rate at high transmit power needs to be found.

\appendices
\section{Approximations for $p_i\ll P$}
\label{FirstApprox}
Since $p_i<P$, we can approximate $\frac{1}{P-p_i}$ in $(\ref{BeamFormingVector})$ as $\frac{1}{P}\left(1+\frac{p_i}{P}\right)$ using Taylor series to the first order.  
Moreover, using Taylor series, the square root term in $(\ref{sqT})$ can be also  approximated as $$\sqrt{\frac{\rho_iP_R}{P}\|\vec{h}_{Rj}\|^2-h_{ij}^2}+\frac{\frac{\rho_iP_R}{P}\|\vec{h}_{Rj}\|^2}{2P\sqrt{\frac{\rho_iP_R}{P}\|\vec{h}_{Rj}\|^2-h_{ij}^2}}p_i.$$
\begin{remark}
Note that this approximation is precise only when $p_i\ll P$, in our case, we only know that $p_i<P$, so this is a rough approximation.
\end{remark}
After substituting in $(\ref{BeamFormingVector})$ and $(\ref{CandEofp1p2})$, we get the following expressions for $f_{11}$ and $f_{22}$
\begin{eqnarray*}
f_{11}&\approx& \mu_{11}(n_1,\rho_1)+\nu_{11}(n_1,\rho_1)\frac{p_1}{P},\nonumber\\
f_{22}&\approx& \mu_{22}(n_2,\rho_2)+\nu_{22}(n_2,\rho_2)\frac{p_2}{P},\nonumber
\end{eqnarray*}
where
\begin{eqnarray*}
% \begin{alignat}{l}
\mu_{11}(n_1,\rho_1)&=&h_{11}-\frac{h_{R1,2}h_{12}}{h_{R2,2}}\\
&&+n_1\frac{\det(\mat{H})}{\|\vec{h}_{R2}\|^2}\left(S_1-n_1\frac{h_{12}h_{R2,1}}{h_{R2,2}}\right),\\
\nu_{11}(n_1,\rho_1)&=&n_1\frac{\rho_1P_R\det(\mat{H})}{2PS_1},\\ 
\mu_{22}(n_2,\rho_2)&=&h_{22}-\frac{h_{R2,2}h_{21}}{h_{R1,2}}\\
&&+n_2\frac{\det(\mat{H})}{\|\vec{h}_{R1}\|^2}\left(-S_2+n_2\frac{h_{21}h_{R1,1}}{h_{R1,2}}\right),\\
\nu_{22}(n_2,\rho_2)&=&-n_2\frac{\rho_2P_R\det(\mat{H})}{2PS_2},
% \end{alignat}
\end{eqnarray*}
with $\mat{H}=[\vec{h}_{R1}\ \vec{h}_{R2}]$, $S_i=\sqrt{\frac{\rho_iP_R}{P}\|\vec{h}_{Rj}\|^2-h_{ij}^2}$, $i\neq j$, $i,j\in\{1,2\}$.
\section{Low $P$ approximations}
\label{SecondApprox}
In the following, we state the proof of Proposition $\ref{RateRegion}$. We consider low $P$, i.e. $P\to0$, and since $p_i<P$, it follows that $p_i\to0$, $i\in\{1,2\}$.
Equations $(\ref{SDMAR1})$ and $(\ref{SDMAR2})$ can be respectively approximated at low $P$ as
\begin{eqnarray}
\label{RindApprox1}
R_{1}^{MAC}&\approx&\frac{\|\vec{g}_{1R}\|^2p_1}{\ln(2)}=r_{1}^{MAC},\\
R_{2}^{MAC}&\approx&\frac{\|\vec{g}_{2R}\|^2p_2}{\ln(2)}=r_{2}^{MAC}.\nonumber
\end{eqnarray}
Equation $(\ref{SDMASumRate})$ can be re-written as
\begin{equation}
\label{SumRateF1p}
R_{sum}^{MAC}= \log(\alpha p_1p_2+\beta p_1+\gamma p_2+1),
\end{equation}
where $$\alpha=\|g_{11}\|^2\|g_{22}\|^2+\|g_{21}\|^2\|g_{12}\|^2-g_{12}g_{21}g_{11}^*g_{22}^*-g_{11}g_{22}g_{12}^*g_{21}^*,$$ $$\beta=\|g_{11}\|^2+\|g_{12}\|^2=\|\vec{g}_{1R}\|^2,$$
$$\gamma=\|g_{21}\|^2+\|g_{22}\|^2=\|\vec{g}_{2R}\|^2,$$
and this can be approximated at low $P$ as
\begin{equation}
\label{RSumApprox1}
R_{sum}^{MAC}\approx\frac{\|\vec{g}_{1R}\|^2 p_1+\|\vec{g}_{2R}\|^2 p_2}{\ln(2)}.
\end{equation}
Notice that the bound $R_{sum}^{MAC}$ is redundant and needs not to be considered for low $P$. Now, equations $(\ref{TreatingInterferenceAsNoiseRates1})$ and $(\ref{TreatingInterferenceAsNoiseRates2})$ can be approximated as
\begin{eqnarray*}
\label{RindApprox2}
R_{1}^{IC}&\approx&\frac{1}{\ln(2)}(\|\mu_{11}\|^2P+(2\Re(\mu_{11}\nu_{11}^*)-\|\mu_{11}\|^2)p_1\nonumber\\
&&-2\Re(\mu_{11}\nu_{11}^*)p_1^2/P)=r_{1}^{IC},\nonumber\\
R_{2}^{IC}&\approx&\frac{1}{\ln(2)}(\|\mu_{22}\|^2P+(2\Re(\mu_{22}\nu_{22}^*)-\|\mu_{22}\|^2)p_2\nonumber\\
&&-2\Re(\mu_{22}\nu_{22}^*)p_2^2/P)=r_{2}^{IC}.
\end{eqnarray*}
As a result of $(\ref{RindApprox1})$ and $(\ref{RindApprox2})$ we can write the rate region at low $P$ as
\begin{eqnarray*}
R_1&\leq&\min(r_{1}^{MAC},r_{1}^{IC}),\nonumber\\
R_2&\leq&\min(r_{2}^{MAC},r_{2}^{IC}),\nonumber\\
% R_1+R_2&\leq&\frac{\|\vec{g}_{1R}\|^2 p_1+\|\vec{g}_{2R}\|^2 p_2}{\ln(2)}.
\end{eqnarray*}
In order to maximize this rate region, we would like to choose a power allocation that maximizes $\min(r_{1}^{MAC},r_{1}^{IC})$ and $\min(r_{2}^{MAC},r_{2}^{IC})$ over $p_1$ and $p_2$ respectively. Since $r_{1}^{MAC}-r_{1}^{IC}$ is a quadratic function of $p_1$, and $r_{1}^{MAC}- r_{1}^{IC}<0$ for $p_1=0$, $r_{1}^{MAC}- r_{1}^{IC}>0$ for $p_1=P$, then $r_{1}^{MAC}-r_{1}^{IC}=0$ admits a solution $\hat{p}_1\in[0,P]$. Similarly, $r_{2}^{MAC}-r_{2}^{IC}=0$ admits a solution $\hat{p}_2\in[0,P]$. After solving the resulting quadratic equations, we get
\begin{eqnarray*}
\hat{p}_1=\frac{\lambda_1+\sqrt{\lambda_1^2+8\|\mu_{11}\|^2\Re(\mu_{11}\nu_{11}^*)}}{4\Re(\mu_{11}\nu_{11}^*)}P,\\
\hat{p}_2=\frac{\lambda_2+\sqrt{\lambda_2^2+8\|\mu_{22}\|^2\Re(\mu_{22}\nu_{22}^*)}}{4\Re(\mu_{22}\nu_{22}^*)}P,\\
\end{eqnarray*}
with $\lambda_1=2\Re(\mu_{11}\nu_{11}^*)-\|\mu_{11}\|^2-\|\vec{g}_{1R}\|^2$ and $\lambda_2=2\Re(\mu_{22}\nu_{22}^*)-\|\mu_{22}\|^2-\|\vec{g}_{2R}\|^2$. Substituting these powers in $(\ref{RindApprox1})$
%  and $(\ref{RSumApprox1})$ 
gives us the rate region achievable by this scheme at low $P$.

\end{document}